\documentclass[]{spie}  %>>> use for US letter paper
\usepackage{booktabs}
\usepackage[]{graphicx}
\title{The LAUE project and its main results} 
\author{E.~Virgilli\supit{a}, F.~Frontera\supit{a,b}, V.~Valsan\supit{a,c}, V.~Liccardo\supit{a,c}, 
V.~Carassiti\supit{d}, S.~Squerzanti\supit{d}, M.~Statera\supit{a,d}, M.~Parise\supit{d}, S.~Chiozzi\supit{d}, 
F.~Evangelisti\supit{d}, E.~Caroli\supit{b}, J.~Stephen\supit{b}, N.~Auricchio\supit{b}, S.~Silvestri\supit{b}, 
A. Basili\supit{b}, 
F.~Cassese\supit{e}, L.~Recanatesi\supit{e}, 
V.~Guidi\supit{a}, V.~Bellucci\supit{a}, R.~Camattari\supit{a},
C.~Ferrari\supit{f}, A.~Zappettini\supit{f}, E.~Buffagni\supit{f}, E.~Bonnini\supit{f},
M.~Pecora\supit{g}, 
S.~Mottini\supit{h} and
B.~Negri\supit{i}
\skiplinehalf
\supit{a} \small\textit{Physics and Earth Sciences Department, University of Ferrara, via Saragat, 1 - 44122 Italy};\\
\supit{b} \small\textit{IASF-INAF via P.Gobetti, Bologna - Italy};\\
\supit{c} \small\textit{Universit\'e de Nice Sophia-Antipolis, Parc Valrose, 06108 Nice Cedex 2, France};\\
\supit{d} \small\textit{INFN sezione di Ferrara,  via Saragat, 1 - 44122 Italy};\\
\supit{e} \small\textit{DTM, Modena, Via Tacito, I-41100 Modena - Italy};\\
\supit{f} \small\textit{IMEM - Parco Area delle Scienze 37/A - 43124 Parma - Ita ly};\\
\supit{g} \small\textit{Thales Alenia Space, Via Enrico Mattei 1, 20064 Gorgonzola, Milan - Italy};\\
\supit{h} \small\textit{Thales Alenia Space, Strada Antica di Collegno 253, 10146 Turin - Italy};\\
\supit{i} \small\textit{ASI, Agenzia Spaziale Italiana, Viale Liegi 26, I-00198 Roma - Italy}.
}

\authorinfo{E-mail to: virgilli@fe.infn.it}

\begin{document} 
\maketitle

\begin{abstract}

 We will describe the LAUE project, supported by the Italian Space 
Agency, whose aim is to demonstrate the 
capability to build a focusing optics in the hard X-/soft gamma-ray domain (80--600 keV). To show  the lens feasibility, the assembling of a Laue 
lens petal prototype  with 20 m focal length is ongoing. Indeed, a  feasibility study, within the LAUE project, has demonstrated that a Laue lens made of petals is feasible.  
Our goal is a lens in the 80-600 keV energy band. In addition to a detailed description of the new LARIX facility, 
in which the lens is being assembled, we will report the results of the project obtained so far.

\end{abstract}


\section{INTRODUCTION}
\label{sec:intro}

Very sensitive observations of celestial hard X--/soft 
gamma--ray photons ($>$80--100 keV) are of key importance 
for the knowledge of many physical processes  in the Universe\cite{Frontera13}~. Unfortunately, till now observations
 beyond 80 keV have been performed with sky 
direct-viewing instruments the sensitivity of which is background limited and  limited is also their angular 
resolution (about $\sim 15$ arcmin in the case of INTEGRAL/IBIS mask telescope).  
To improve flux sensitivity and angular resolution, the use of gamma-ray focusing optics is crucial.
The LAUE project, supported by the Italian Space Agency (ASI), is a follow-up of the HAXTEL project
and is devoted to create a technology to build a Laue lens with long focal length (20--100 m) able 
to focus photons in the 80--600 keV energy range. In the HAXTEL project, two prototypes of Laue lens 
with 6 m focal length (FL) have been built using flat mosaic copper crystals. 
The experience acquired from the HAXTEL 
project has shown that, for FL longer than 10--15 m a new assembling technique is needed. 

Moreover, a new generation of crystals must be developed for Laue lenses, given that the Point Spread Function (PSF) of a flat crystal
is related to the size of the crystal itself. Flat crystals are also limited in terms of diffraction efficiency given that 
their theoretical limit is the 50\%. In the last years, new technologies have been developed to obtain 
bent crystals, whose properties surpass those shown by flat mosaic crystals
in terms of efficiency and focusing capability.
                                                                                                                                                                                                            
\begin{figure}[!h]
\begin{center}
\includegraphics[scale=1.1]{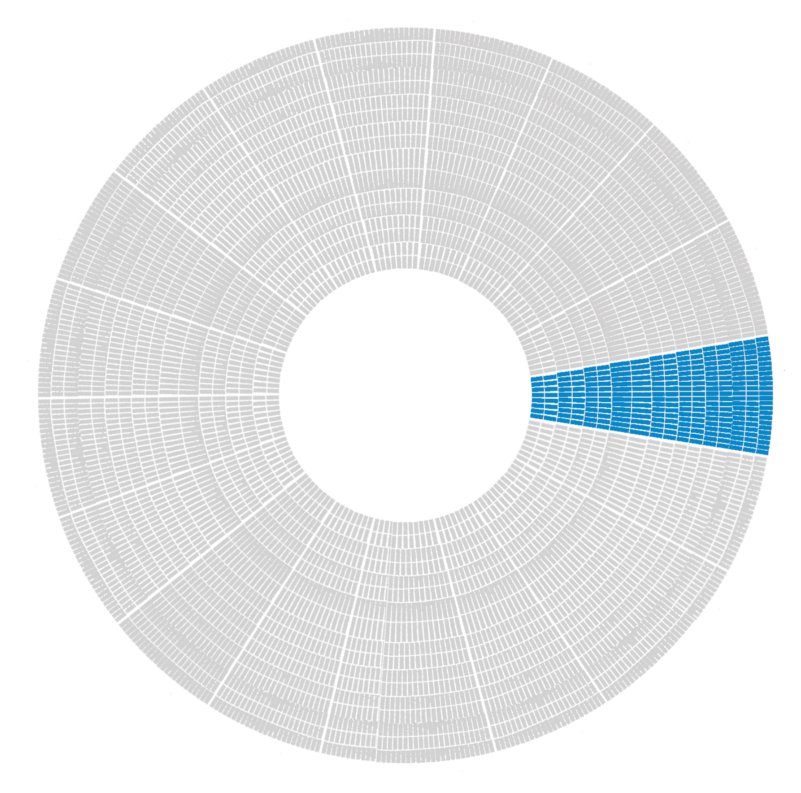}
\caption{\footnotesize Sketch showing the concept of modularity of a Laue lens. The lens is made of crystals placed in concentric 
rings, from the inner part (diffracting the higher energies) to the outer radius (lower energies). In the LAUE project, 
a petal as part of a whole lens will be built and test (blue part).}
\label{fig:completelens} 
\end{center}
\end{figure}

In the LAUE project, an industrial study has been performed to check the feasibility of a modular lens. An entire 
lens is made of petals. Each petal can be made of sub-petals (or petal sectors). The crystals tiles are accommodated in 
the petal sectors (see Fig.~\ref{fig:completelens}).

\begin{figure}[!h]
\begin{center}
\includegraphics[scale=0.35]{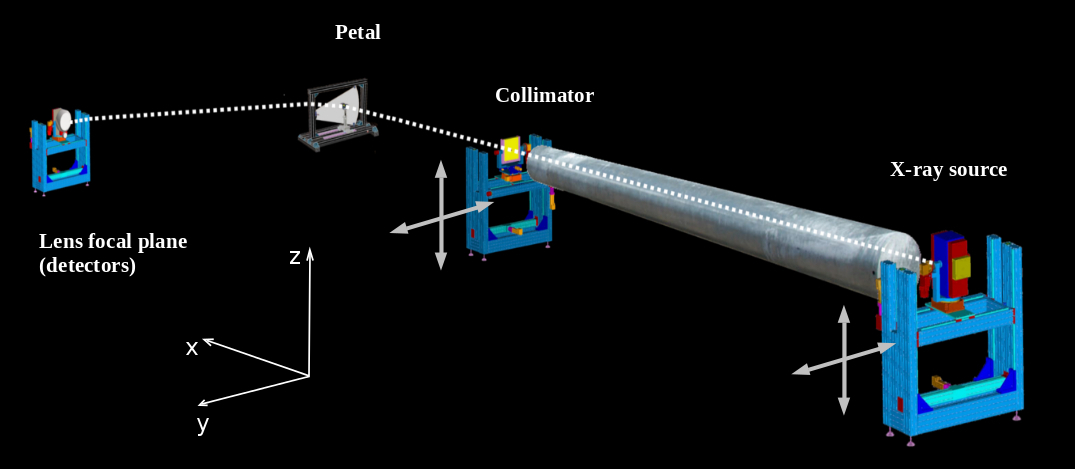}
\caption{\footnotesize Layout of the LAUE apparatus that is on going of realisation in the facility of the 
University of Ferrara.}
\label{fig:tunnel} 
\end{center}
\end{figure}

One of these petals is being built as a demonstration of the project success. 
The adopted technology consists in the positioning of the crystal tiles on the lens frame
under the control of a gamma-ray beam. The lens frame is kept fixed during the entire lens assembling process, 
while the gamma-ray source and a mechanical collimator, which includes a slit with variable aperture, are moved together 
along the Y and Z axes to simulate a parallel beam.
Each diffracting crystal is correctly translated and oriented in order to focus the beam photons on the lens focal plane. 
The final crystal tile position is then glued  upon the lens frame. The main elements of the project 
are summarised in Fig.~\ref{fig:tunnel} where also the reference axes are shown to clarify the 
description of each component and its possible motion.

\section{The LARIX facility}

The entire apparatus for both assembling and testing the lens is installed in the LArge Italian X-ray laboratories (LARIX) located 
at the Physics and Earth Science Department of the University of Ferrara. The laboratories include an experimental room (LARIX A) with a 12~m 
long facility and a 100~m long tunnel.
The facility in LARIX A has been successfully used for calibrating JEM~X aboard the INTEGRAL satellite 
\cite{Loffredo03}~, and building and testing two prototypes of Laue lenses with 6 m focal length made of 20 copper flat mosaic crystals 
\cite{Frontera08, Virgilli11}~. 
%The crystal were 
%placed on a carbon fiber support in a circular ring with 18 %cm radius, and the prototypes were conceived for 
%refocusing (6 m focal length) the 100 keV photons coming 
%from a source placed at 6 m far from the lens.

\begin{figure}[!h]
\begin{center}
\includegraphics[scale=0.42]{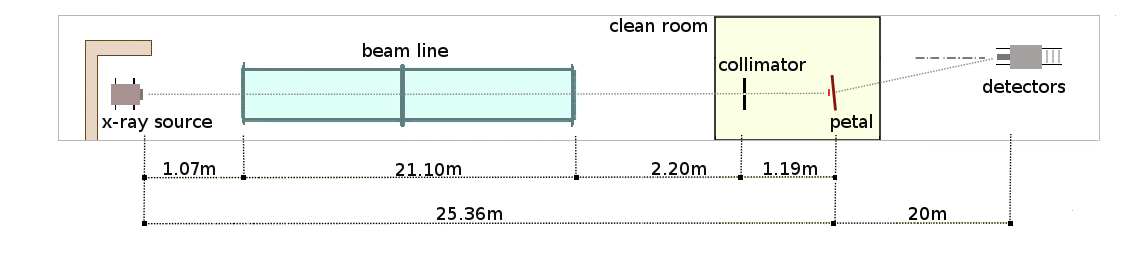}
\caption{\footnotesize Sketch of the tunnel with the relative distances between the sub-systems.}
\label{fig:sketch} 
\end{center}
\end{figure}

The same facility is supporting the LAUE project and is used 
for a preliminary qualification of each crystal tile in terms of reflectivity and curvature radius before to be used for
the LAUE project. The 100 m long tunnel is instead the ideal place for the LAUE project. 
It allows to obtain a small divergence of the beam impinging on each crystal
and to build Laue lenses with long focal length. In Figure~\ref{fig:sketch} it is shown a sketch of 
the tunnel and the relative distances between the installed sub-systems.

The complete apparatus mounted in the LARIX facility consists of the following sub-systems that will be 
in detail described in the next subsections.

\begin{itemize}
 \item X-ray source;
 \item Beam-line with vacuum environment;
 \item Clean room with thermal and humidity control;
 \item Adjustable mechanical slit;
 \item Hexapod positioning device and crystal holder;
 \item Petal frame; 
 \item Focal plane detectors;
 \item Translation and rotation systems for the motion of each subsystem (source, collimator slit, crystals);
\item Rail for translating the focal plane detectors along the lens axis;
 \item Hardware and software needed for the remote control (Ground Support Equipment, GSE). 
\end{itemize}

\subsection{The X-ray sources} 

The tunnel of the LARIX facility is equipped with two X-- and gamma--ray generators. The first is a portable 
betatron that provides a photon source
with maximum energy of 2.5 MeV. Thanks to the 
extremely broad energy band, the betatron could be exploited for assembling and testing Laue lenses with nominal 
passband of 80--600 keV, which is our final astrophysical goal. 

\begin{figure}[!h]
\begin{center}
\includegraphics[scale=0.28]{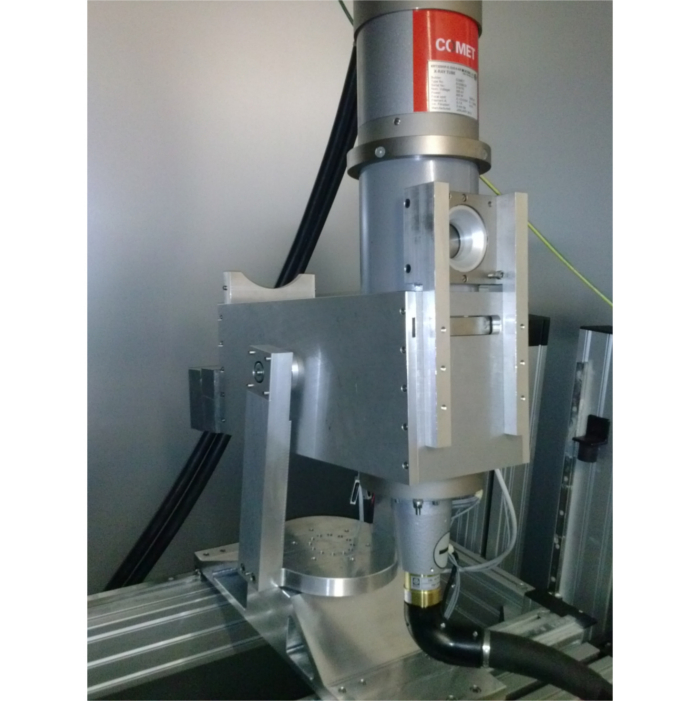}
\includegraphics[scale=0.28]{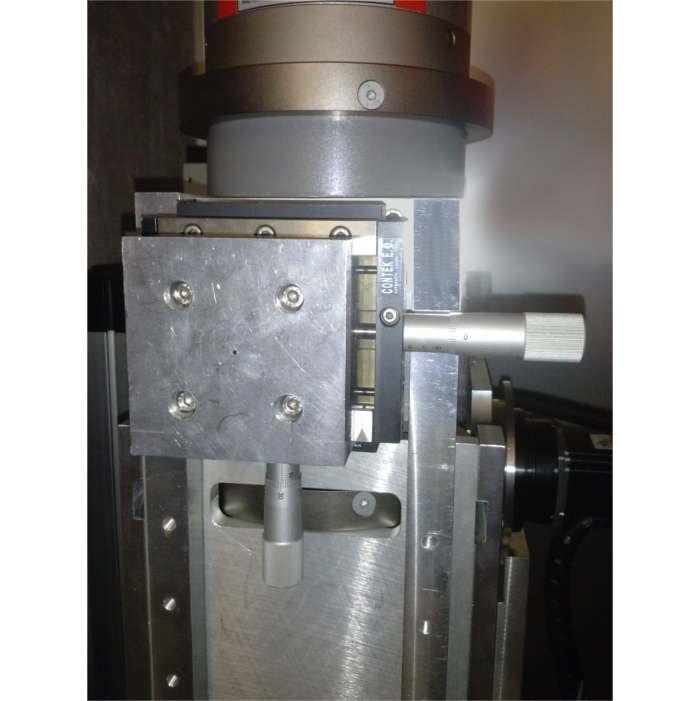}
\caption{\footnotesize The X-ray source that is being used for assembling and testing the lens petal ({\it left}). 
and a detail showing the 1 mm lead collimator placed in front of the X-ray tube aperture ({\it right}).}
\label{fig:source} 
\end{center}
\end{figure}

The second generator makes use of a traditional X--ray tube (provided by Bosello Technology) with a fine focus 
of 0.3 $\times$ 0.4 mm$^2$, a maximum voltage of 320 kV, and a maximum power of $\sim$ 1800 W(Fig.~\ref{fig:source}~{\it left}).
The advantage of this gamma--ray generator is that is much brighter than the betatron. 
Thus, to speed up the assembling phase of the lens petal prototype foreseen within the LAUE project we 
have adopted the latter generator. This choice has limited the upper threshold of energy band of the lens 
petal prototype to $\sim 300$~keV.

\begin{figure}[!h]
\begin{center}
\includegraphics[scale=0.65]{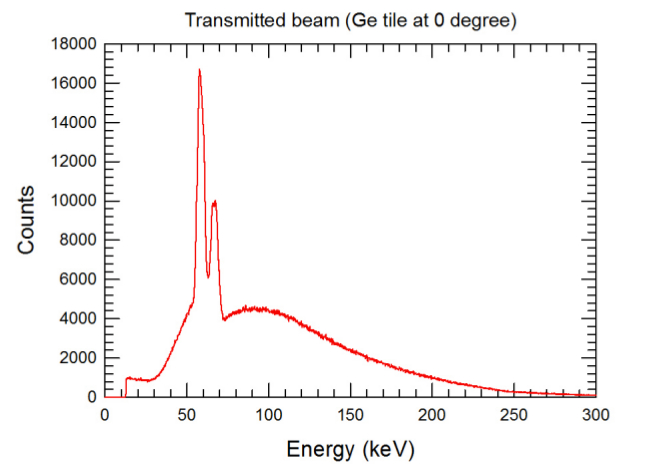}
\caption{\footnotesize Spectrum of the transmitted beam through a Germanium (111) crystal tile. The collimator size 
close to the target is 0.5 $\times$ 6 mm$^2$. The X-ray generator parameters are HV= 250 V and current = 2.4 mA.}
\label{fig:directbeam} 
\end{center}
\end{figure}

In table~\ref{tab:tubo} we  report the main properties of the adopted X-ray generator. A typical beam 
measured by the spectrometer coming from the source and absorbed by a 2 mm thick Germanium crystal is shown in Fig.~\ref{fig:directbeam}.

A 20 mm thick Tungsten shield with a 3 mm diameter hole has been installed in front of the aperture of 
the X-ray tube to reduce the X-ray beam. A further collimation is obtained with a drilled (1 mm diameter) 
50 mm thick lead collimator (Fig.~\ref{fig:source}~{\it right}).
Source plus collimators can be moved on the plane Y-Z (X is the beam axis). Thanks to the overall collimation, the background level at the 
lens focus has been highly reduced.

\begin{table}[!h]
\caption{\footnotesize Main properties of the X-ray tube that is being used for the assembling phase.}
\begin{center}
\begin{tabular}{ll}
\toprule
Nominal Tube Voltage (kV) & 320 \\
Maximum power  (W)  & 800 / 1800 \\
Focal spot size (mm)  & d$_{small}$ = 0.4\\
                      & d$_{big}$  = 1.0\\
Inherent filtration    &  3 mm Be\\
Target material      &   W\\
Cooling medium  &  Oil\\
Weight (kg) & 40\\
\hline
\end{tabular}
\end{center}
\label{tab:tubo}
\end{table}

\subsection{Beam-line and clean room} 

\begin{figure}[h!]
\begin{center}
\includegraphics[scale=0.123]{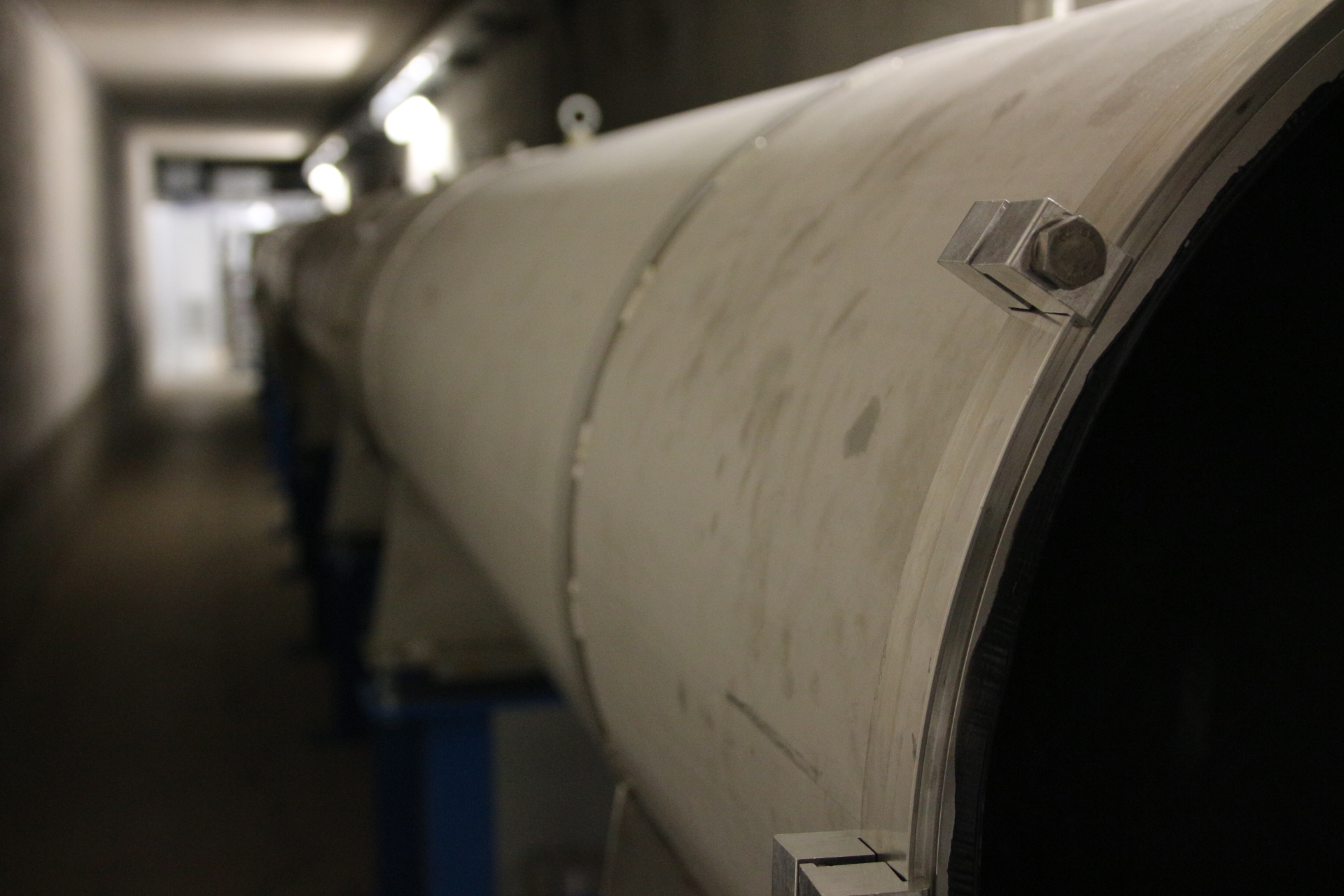}
\includegraphics[scale=0.89]{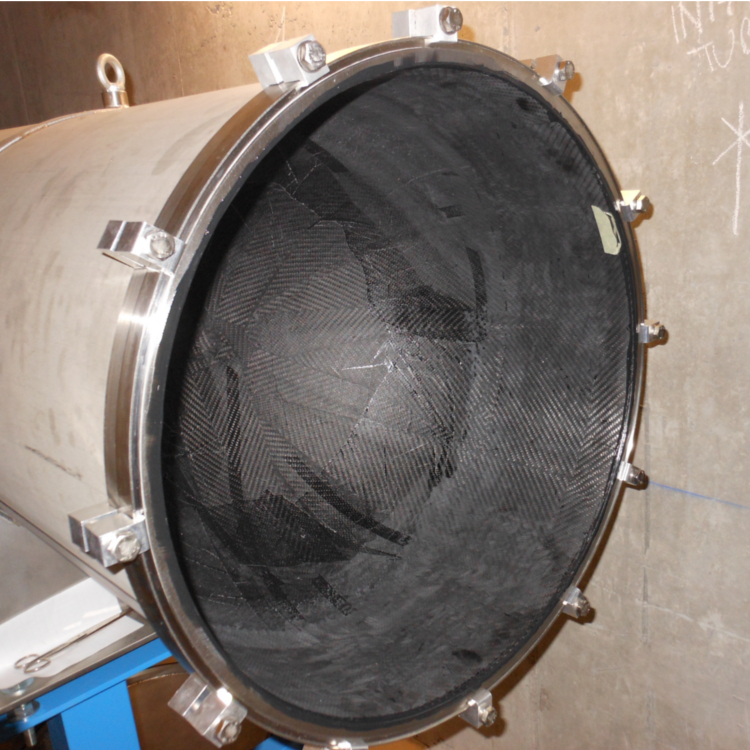}
\caption{\footnotesize The 21 m beam line ({\it left}) in which is carried out the vacuum for avoiding absorption and scattering. The 
carbon fibre windows ({\it right}) ensure the system to be completely sealed at the working pressure.}
\label{fig:beamline} 
\end{center}
\end{figure}

After the collimated photon beam pass through a 21 m long under--vacuum beamline to avoid absorption and 
scattering of the generated beam off air. The beamline is made of a set of 7 modules. Each module is a tube of stainless steel with a length of 3 m. The vacuum environment is guaranteed by three vacuum pumps that keep stable the pressure below  1 mbar. The pressure read out control is performed with a set of probes, one for each
pump. The X--ray entrance and exit windows of the beamline are made of carbon 
fiber 3 mm thick which guarantee a gamma--ray transparency   $>90$\% @ 100 keV.
Figure \ref{fig:beamline} shows a detail of the beam-line and one of the carbon fiber entrance windows. 

Given the requested positioning accuracy of the single crystal tiles on the lens petal frame,  lens petal prototype is assembled in a low-laminar-flow clean room (dimensions 2 $\times$ 8 m$^2$, class 10$^5$, US FED STD 
209E Clean room Standards). It is equipped with a thermal control (within $1\,^{\circ}{\rm C}$ accuracy) and an 
hygrometric control (relative humidity $\Phi$ = 50\% within an error of 10\%).

\subsection{Mechanical and motorized  collimator} 

After the beam-line the beam impinges on a collimator equipped with a slit with variable aperture. It has to be moved together with the X-ray generator along the Y-Z axes to artificially reproduce
the presence of a source coming from the infinity. Thus X-ray beam so obtained is always parallel to itself and to the lens axis while it is impinging on every crystal tile during the crystal assembling phase. 
Thus the collimator is equipped with motors that can translate it in the X-Y plane and it is provided with three motors to rotate the slit around the X, Y and Z axes.

\begin{figure}[h!]
\begin{center}
\includegraphics[scale=0.165]{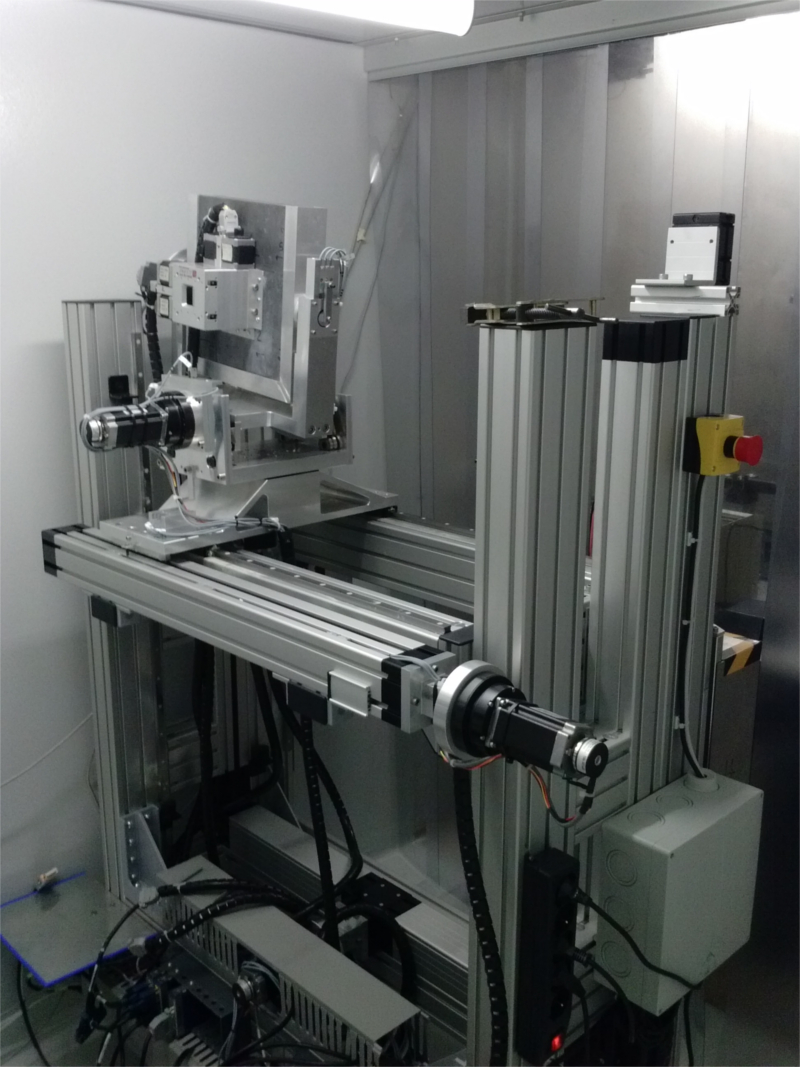}
\includegraphics[scale=0.22]{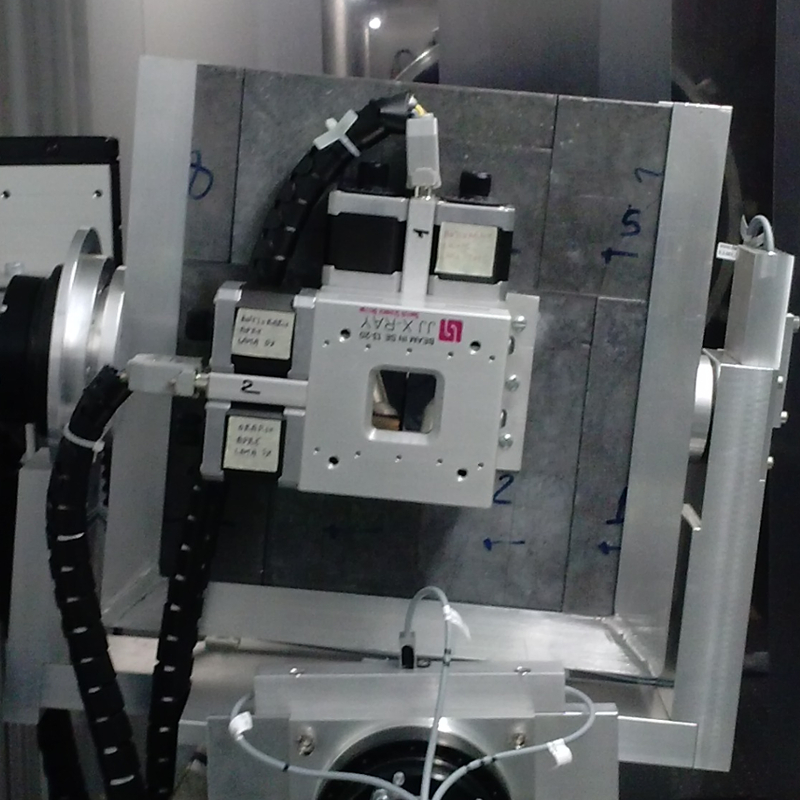}
\caption{\footnotesize Collimator and carriage ({\it left}) and a detail of the motorised collimator mounted on 
the 200 $\times$ 200 mm$^2$ lead panel ({\it right}).}
\label{fig:JJ} 
\end{center}
\end{figure}

The collimator has also another important role. It has to shield the beam coming from the source, except that around the beam axis. The shielding is achieved by a 
lead panel with 200 $\times$ 200 mm$^2$ cross section and 50 mm thick. In the center of the panel it has a 30 $\times$ 30 mm$^2$ aperture window. In correspondence of this window a motorized slit with variable aperture is located. The variable aperture is obtained by means of four crossed and independent 20 mm thick blades made of Tungsten Carbide 
(density 15.6 g/cm$^3$, 1 \% transparency @ 600 keV).
In Fig.~\ref{fig:JJ}~{\it left} it is visible the collimator slit and its carriage while a detail showing the slit located on the 200 $\times$ 200 mm$^2$ lead screen is given in Fig.~\ref{fig:JJ}~{\it right}. 

\subsection{The hexapod system}

\begin{figure}[!h]
\begin{center}
\includegraphics[scale=0.15]{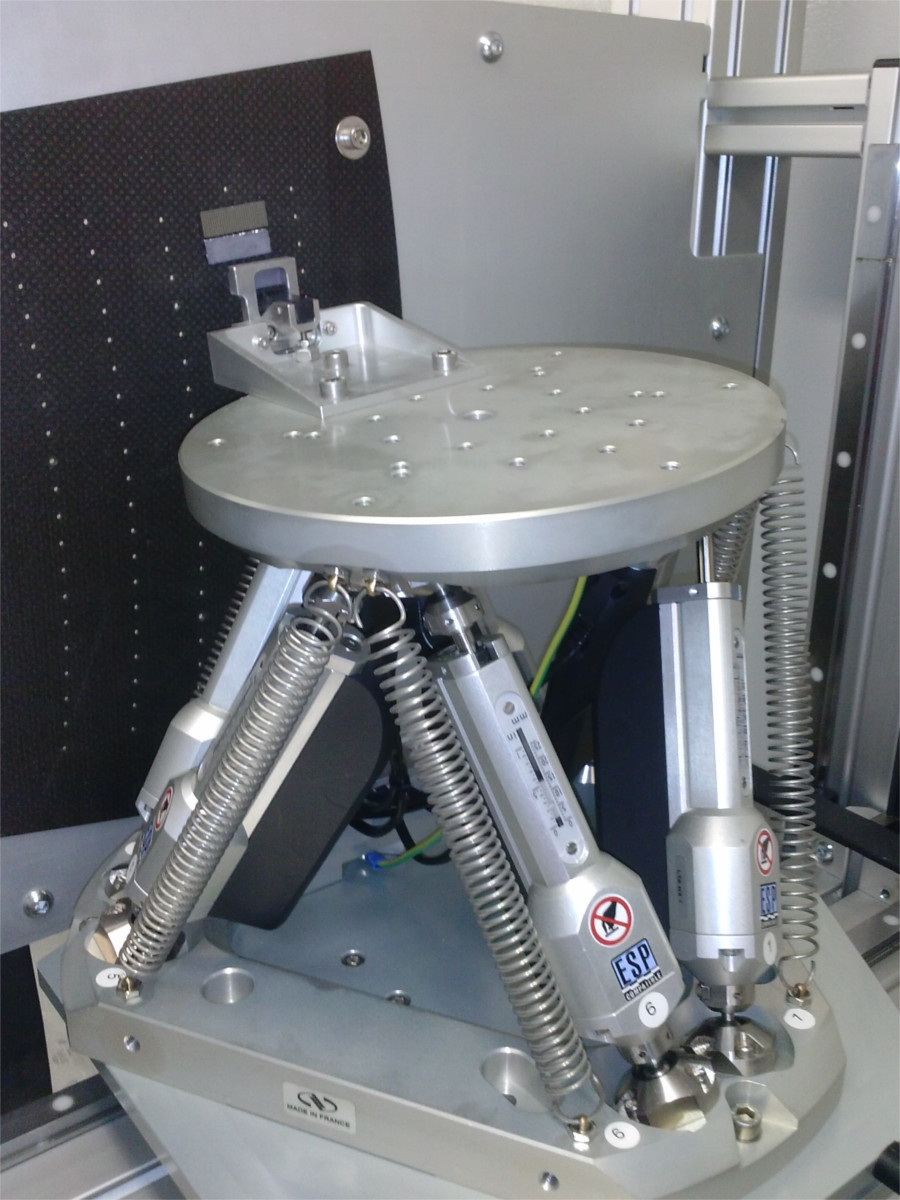}
\includegraphics[scale=0.195]{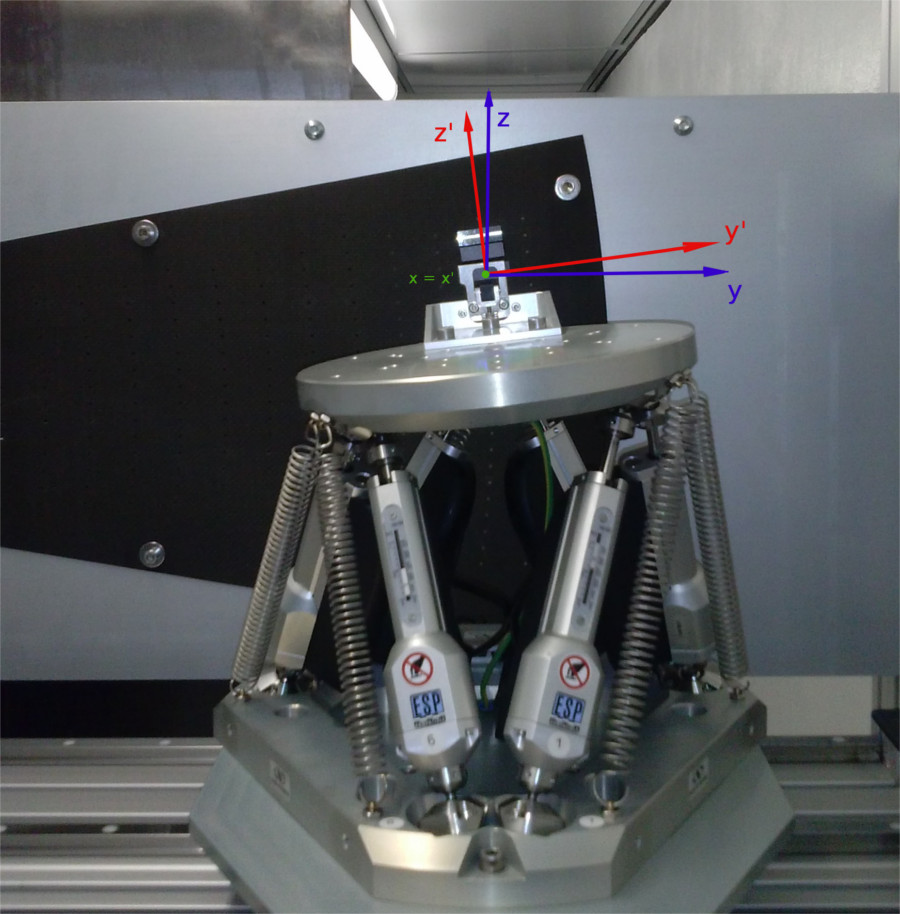}
\caption{\footnotesize The hexapod system that is used to tilt and place each crystal in the proper position for the 
diffration. General view ({\it left}) of the hexapod and its back view ({\it right}). The reference axes Y, Z and the 
principal axes of rotation Y', Z' are also reported, referred to a generic crystal slot (position \#~25).}
\label{fig:exa} 
\end{center}
\end{figure}

The orientation and the positioning of each crystal on the carbon fiber petal frame is performed with a 6-axis 
hexapod system (Fig.~\ref{fig:exa}). The rotation and the linear motion uncertainties are 1 arcsecond and 1 $\mu$m, respectively.
While the coarse translations along the Y and Z axis are provided by the carriage, the hexapod is mainly used for the three 
rotations. The first is the rotation around the beam axis (X) to place each crystal at the nominal position. The other two 
rotations around Z' and Y' axis of the crystal rest frame), which
in general do not coincide with the laboratory reference system Y and Z, allow to find the proper 
diffraction angle. The hexapod is also provided with a further axis (along X) to approach (remove) the crystal  
to (from) the petal frame.

\begin{figure}[!h]
\begin{center}
\includegraphics[scale=0.20]{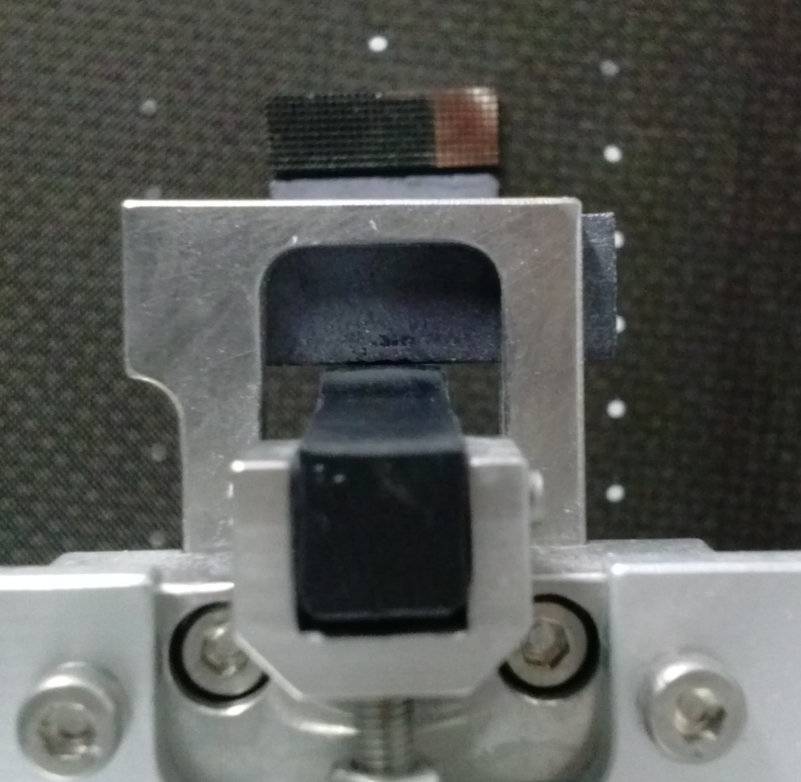}
\includegraphics[scale=0.20]{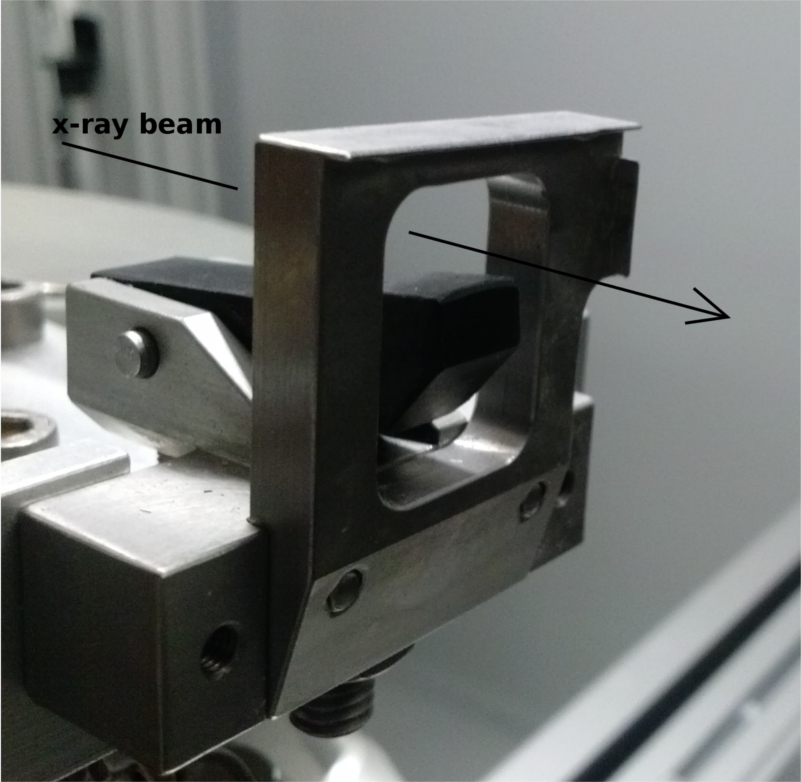}
\caption{\footnotesize Details of the holding system with one crystal made of Gallium Arsenide mounted 
on it ({\it left}). A particular of the anchoring borders and the pivot clamping from the bottom~({\it right}).}
\label{fig:holder} 
\end{center}
\end{figure}

The crystal tile holder has been developed in order to firmly hold the tile and, at the same time, to 
leave free the front and back main faces of the crystal, where the incident beam arrives and the diffracted beam is produced. In order to fill as much as possible the lens frame with crystals, the left
side of the tile is free (see Fig.~\ref{fig:holder}).

\subsection{The petal frame} 

\begin{figure}[!h]
\begin{center}
\includegraphics[scale=0.25]{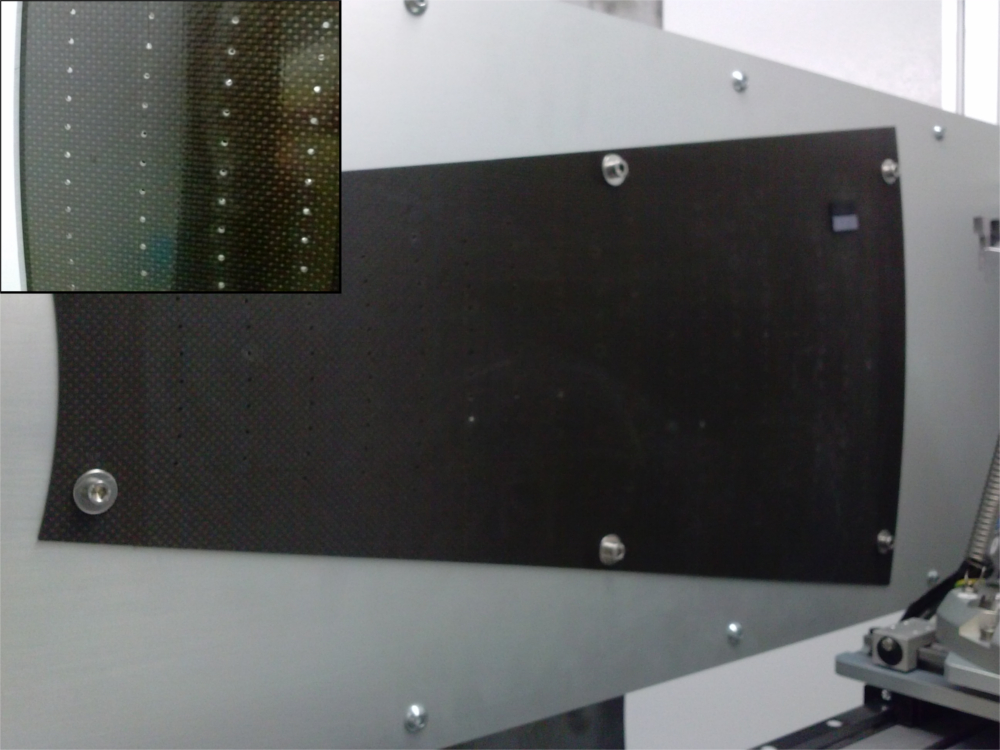}
\caption{\footnotesize The petal frame used as a support for the Laue lens. In the small box is visible a 
small portion of the petal with the holes used to inject the adhesive from the back side.}
\label{fig:petalo} 
\end{center}
\end{figure}

The lens petal frame holds the crystal tiles. The crystals
are properly aligned and placed in the correct position, then glued and fixed on the petal frame. 
Its total thickness (2.3 mm) is realized with a superposition 
of 10 layers of carbon fiber. Assuming a an inter-distance between contiguous crystals of 0.2 mm, we have designed the cell in which each crystal has to be positioned on the frame 
(see Fig.~\ref{fig:petalo}). In the center of each cell,  the frame is drilled, and through the hole the resin is injected for fixing the crystals to the frame. From the location of the cell and the assumed focal length of 20~m, the expected energy in correspondence of the center of each cell is computed.

\begin{figure}[!h]
\begin{center}
\includegraphics[scale=0.7]{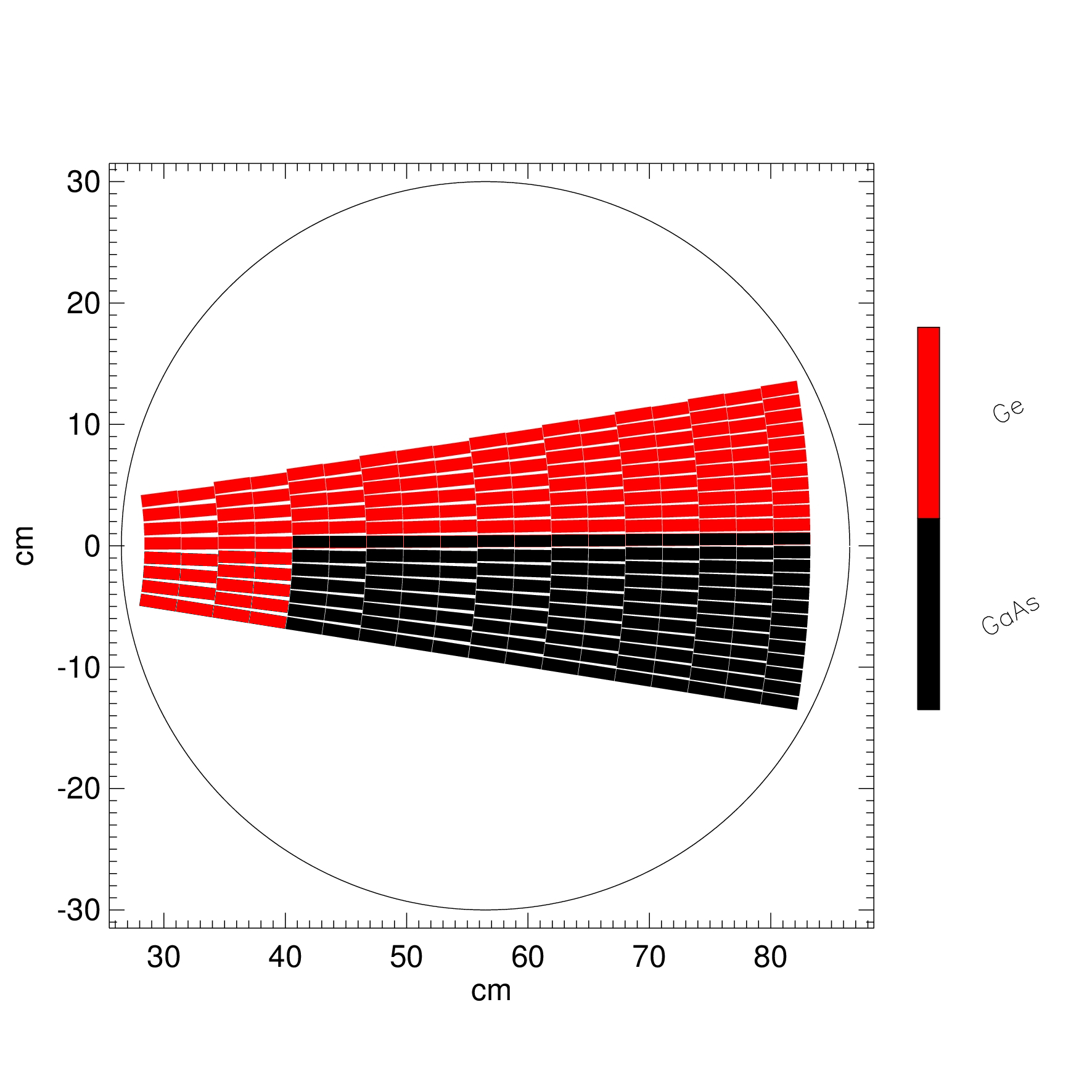}
\caption{\footnotesize Sketch of the lens petal which is being built in the Ferrara LARIX facility. The Ge (111) tiles are accommodated in the red sector, 
while the GaAs (220) crystals are accommodated in the black sector. The 60 cm diameter circle represents the beam-line diameter 
within which the lens petal has to be inscribed.}
\label{fig:petal} 
\end{center}
\end{figure}

In Table~\ref{tab:petal}, the properties of the lens petal prototype, which is being assembled, are  reported. 
The energy passband is defined 
by the inner and outer radius of the lens petal. 
For the designed petal, these values are those allowed by the beamline diameter within which the petal has to be 
inscribed (see Fig.~\ref{fig:petal}). 
The lens passband is also driven by the X-ray source
maximum energy (about 300 keV). The upper part of the petal (red portion 
in Fig.~\ref{fig:petal}) will be filled with Ge (111) tiles while the bottom part will be filled with GaAs (220) tiles. 
Being the atomic d-spacing different for the two specimen, the relative passband is different. 
In the former case the diffracted energy band is 90 -- 267 keV while for the latter is 148 -- 304 keV.

\begin{table}[!h]
\caption{Parameters of the petal that is being build in the LARIX facility.}
    \begin{tabular}{llll}
  \toprule
                                      & GaAs(220) sector              & Ge(111) sector              & Entire petal     \\ 
    \toprule                                  
    Focal length                       & 20 m                           & 20 m                         & 20 m\\
    Energy range                       & 148 -- 304 keV                 & 90 -- 267 keV                & 90 -- 304 keV\\
    No. of rings                       & 14                             & 18                           & 18\\
    Minimum radius                     & 40.66 cm                       & 28.40 cm                     & 28.40 cm\\
    Maximum radius                     & 83.47 cm                       & 83.47 cm                     & 83.47 cm\\
    No. of crystal tiles               & 119                            & 155                          & 274\\
    Crystal dimension (mm$^a$)         & 30 $\times$ 10 $\times$ 2      & 30 $\times$ 10 $\times$ 2    & 30 $\times$ 10 $\times$ 2\\
    Crystal mass (total)               & 2.5 g $\times$ 119 = 297.5 g ~~~~~ & 2.07 g $\times$ 155 = 320.85 g ~~~~~ & 618.35 g\\
    \bottomrule
    \end{tabular}
  \newline
{\small $^a$ radial dimension, tangential dimension, thickness}
    \label{tab:petal}
\end{table}

\subsection{The focal plane detectors} 
\label{sec:laue}

\begin{figure}[h!]
\begin{center}
\includegraphics[scale=0.32]{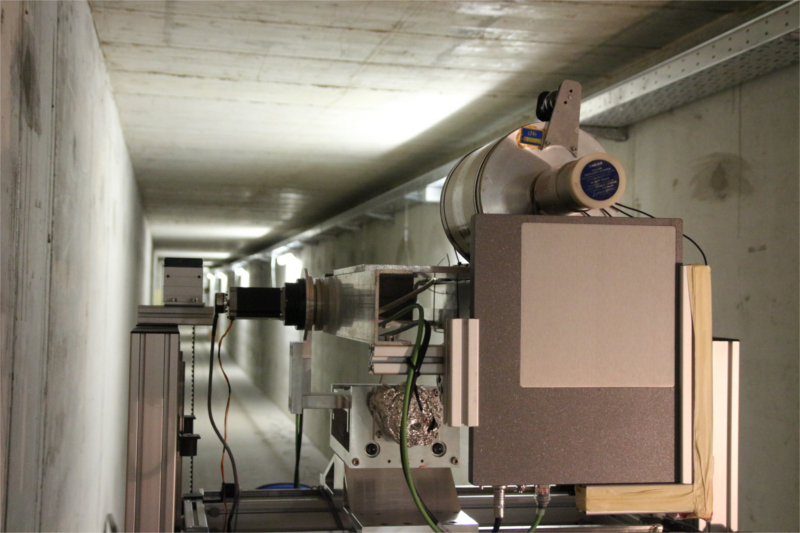}
\caption{\footnotesize The focal plane detectors used in the LAUE project: the flat panel detector is positioned below the HPGe detector.}
\label{fig:detectors} 
\end{center}
\end{figure}

The measurement of the diffracted signal produced by each crystal is performed by means of two focal plane detectors. 
Both are placed on a carriage at the focal distance of 20 m from the petal frame. The carriage can be moved along the lens axis by means of a 15 m long rail. The detectors are shown in Fig.~\ref{fig:detectors}.

The X-ray imaging detector  has a spatial 
resolution of 200 $\mu$m. The main properties of the imager are summarised in Tab.~\ref{tab:imager}. 
The imager is a flat panel 
based on a CsI scintillator (0.8 mm thick) that converts the X and gamma-rays into optical light 
that is afterwards
converted into an electric signal from an array of photodiodes. The imager allows to detect the space distribution of the photons diffracted by the cross section of each crystal which is irradiated by the incident beam. 
By means of an IDL code the Y' and Z' barycenter coordinates of the diffracted beam are estimated and 
converted into two rotation angles in order to properly shift the diffracted image barycenter over the reference pixel 
(the centre of the detector @ px 512-x, 512-y).

\begin{table}[!h]
\caption{Main parameters of the flat panel selected as imager detector for the LAUE project.}
\begin{center}
\begin{tabular}{ll}
\toprule
   Energy range &  40 keV -- 15 MeV\\
   Spatial resolution ($\mu$m) &  200\\
   Active area (cm$^2$)       & 20.48 $\times$ 20.48\\
   Active pixel number     & 1000 $\times$ 1000\\
\bottomrule
\end{tabular}
\newline
\label{tab:imager}
\end{center}
\end{table}

The cooled HPGe spectrometer provides the spectrum of the diffracted photons. The spectrometer allows to establish the correct positioning of each crystal on the lens, given that the centre of each crystal
has to diffract a well defined energy. The detector also allows to have a control of the direct as well as the transmitted
beam through the crystals, to estimate their reflection efficiency. In Table~\ref{tab:spectrometer} are 
reported the main parameters of the detector.

\begin{table}[!h]
   \caption{Main parameters of the spectrometer used for the LAUE project.}
  \begin{center}
    \begin{tabular}{ll}
       \toprule
    Operational energy range &  3 keV -- 1  MeV\\
    Energy resolution @ 122 keV  & 0.4\% \\
    Semiconductor material & P-type High-Purity Germanium\\
    Dewar capacity            &  3 l\\
    Be window thickness (mm)          & 0.254\\
    \bottomrule
    \end{tabular}
    \newline
    \label{tab:spectrometer}
  \end{center}
\end{table}

The detectors are placed on a carriage which allows the motion in the Y-Z plane with an uncertainties of $\pm$ 0.5 mm in 
each direction. It also permits to select which detector has to be positioned on the focus.
In order to allow the measurement 
of both the direct and the diffracted beam, the spectrometer overhangs the carriage using a counterbalance system. The system holding the detectors can also rotate along the three axes.

\subsection{Crystals selection} 
\begin{figure}[!h]
\begin{center}
\includegraphics[scale=0.08]{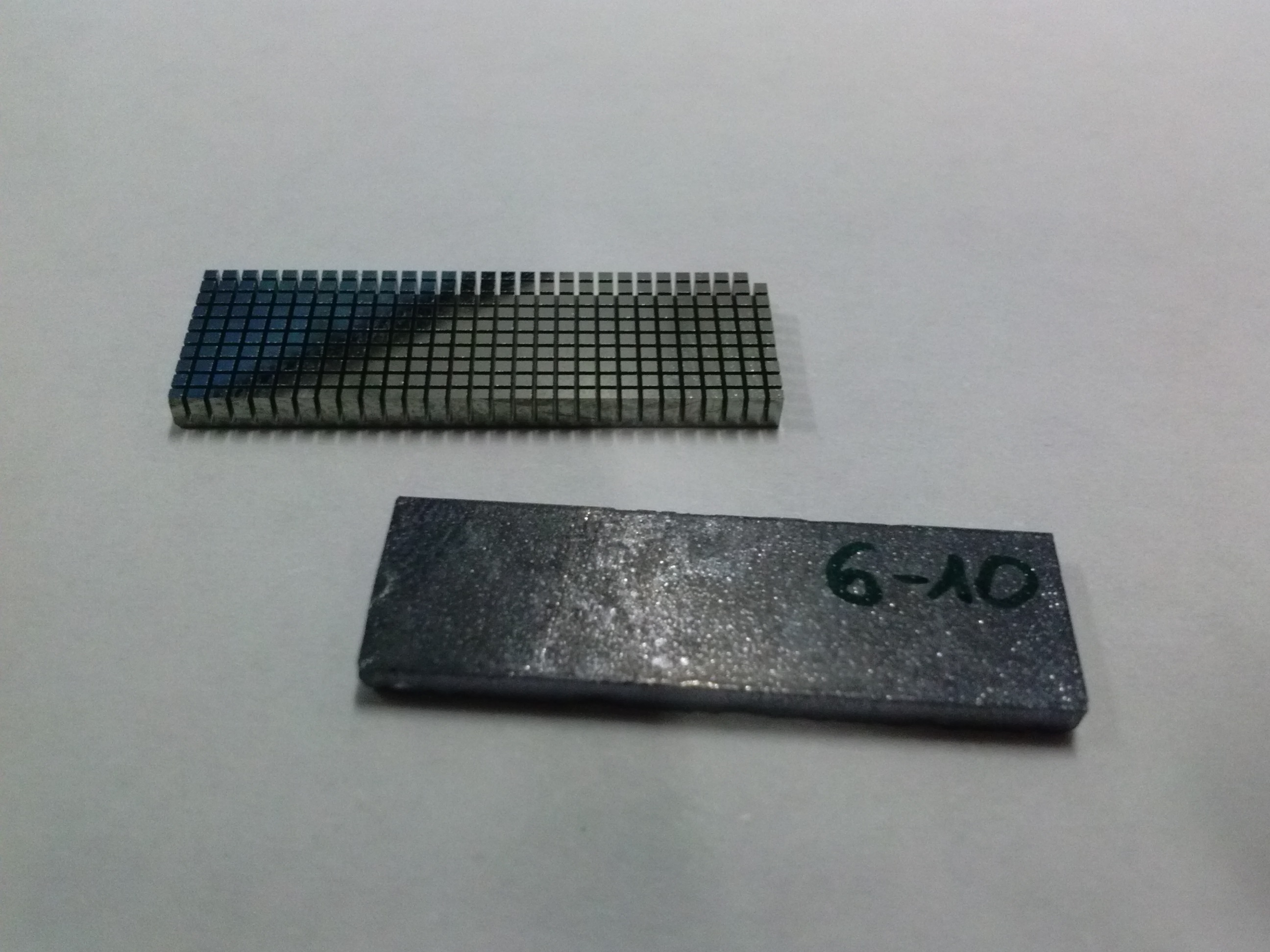}
\caption{\footnotesize Two crystals selected to be placed on the frame of the lens petal. For the Ge (111) tile ($top$) 
it is apparent the set of indentations for  bending the crystal, while the curvature of GaAs (220) crystals ($bottom$) is achieved 
by surface lapping.}
\label{fig:tipocristalli} 
\end{center}
\end{figure}

The entire set of tiles used for the experiment is provided  by two partners of the project: the Sensor and Semiconductor 
Laboratory (LSS) of the University of Ferrara and the CNR/IMEM, Parma.
The lens petal will be assembled  with bent crystal tiles of Germanium (111) (LSS), and Gallium Arsenide (220) (IMEM).
The curvature of the Ge (111) tiles are obtained by indentations~\cite{Bellucci11, Guidi13} while 
the GaAs (220) tiles are bent by surface lapping~\cite{Marchini11, Buffagni11}$.$ For both the specimen an excellent cylindrical 
or spherical shape with the desired curvature radius can be obtained. The proper curvature of Ge (111) tiles is obtained by 
finely tuning the parameters of the process like the grooves number, width and depth of indentation, 
speed of the process~\cite{Camattari13}$.$ For the GaAs (220) crystals, the lapping exposure time and the fine-grained paper 
rules the achieved curvature. An estimation of the curvature radius and reflectivity on a sub-set of 
both materials can be found in \citenum{Liccardo13}. Two samples are shown in Fig.~\ref{fig:tipocristalli}.

Thanks to the focusing capability of each single tile, it has been 
possible to choose the crystal cross section to be 30 $\times$ 10 mm$^{2}$, with the longer 
size radially placed. The focusing effect will make smaller the point spread function (PSF), according to 
what has been estimated from the Monte-Carlo simulations, also taking into account the possible misalignments 
between the tiles, and the curvature uncertainties~\cite{Valsan13}$.$ A big radial dimension reduces the total number of crystals, 
minimizing the error budget due to the misalignment of each crystal during the assembling phase.
The thickness $t$ of the crystal tiles is 2 mm, for each type of crystal material. This  
is a good compromise between the need of a high reflectivity and the current limitation in the thickness imposed
by the bending technology adopted.

\subsection{The Ground Support Equipment}

Each sub-system is equipped with a set of linear and rotational stages of motion:

\begin{itemize}
\item X-ray source: Y and Z traslations, R$_Y$ and R$_Z$ rotations; 
\item Collimator: Y and Z traslations, R$_X$, R$_Y$ and R$_Z$ rotations;
\item Hexapod carriage: Y and Z traslations;
\item Hexapod: 6 d.o.f. (X traslation is allowed manually);
\item Detectors: Y and Z traslations,  R$_X$, R$_Y$ and R$_Z$ rotations (X traslation is allowed manually).
\end{itemize}

\begin{figure}[!h]
\begin{center}
\includegraphics[scale=0.23]{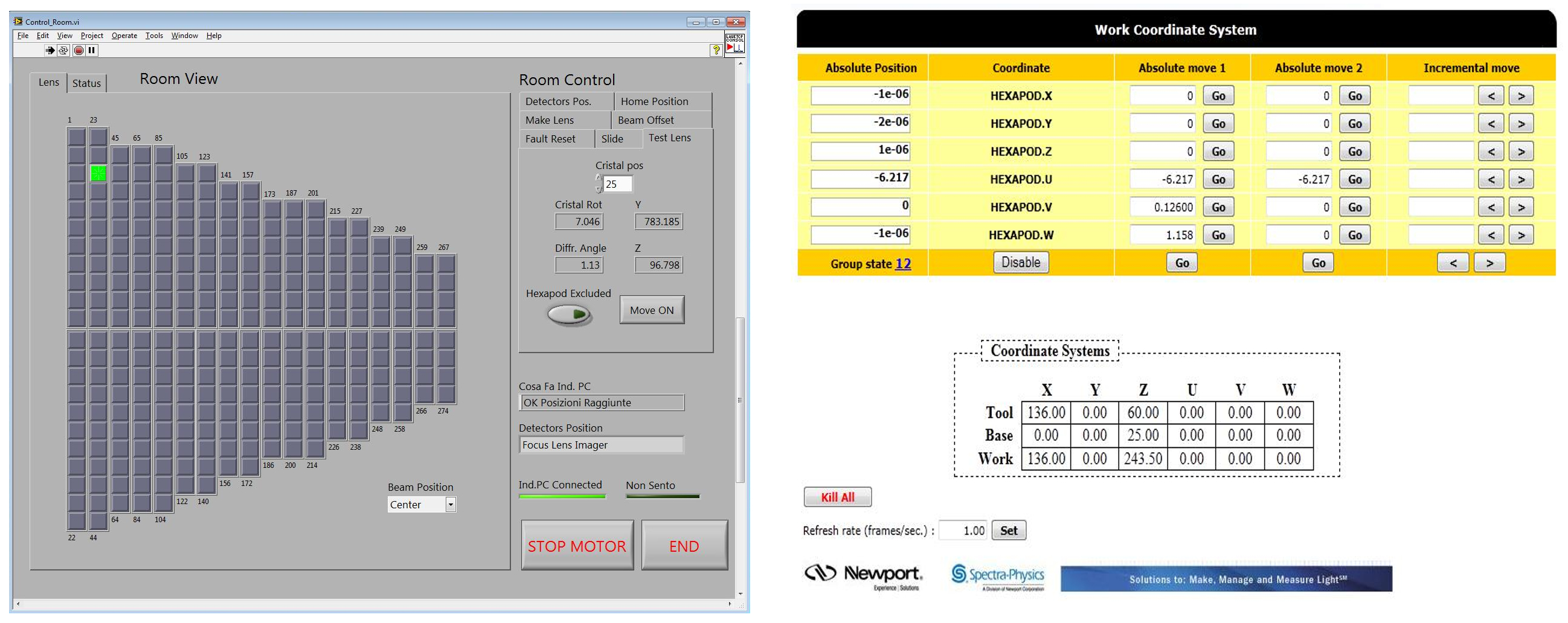}
\caption{\footnotesize Front-end of the LabVIEW management software for the remote control of 
each carriage ({\it left}) and the dedicated software for the hexapod control ({\it right}).}
\label{fig:software} 
\end{center}
\end{figure}

The entire subset of carriages are supposed to be remotely controlled from the control room. 
For the control of each sub-system a proper LabVIEW code has been developed\cite{Caroli13}~. 
The front-end of the remote management suite is shown in 
Fig.~\ref{fig:software}({\it left}). During the assembling procedure as well as in the testing phase a selection of the generic 
light indicators (indicating the crystal to be 
placed at the corresponding lens position) make the X-ray beam (source and collimator slit) and hexapod/crystal holder  
automatically moving to the desired position.  
The control of the hexapod is instead not implemented in the LabVIEW code but is obtained exploiting a
dedicated software (Fig.~\ref{fig:software}({\it right}).

\section{Mechanical, optical and gamma-ray alignments of the facility}

Each module of the beamline has been placed horizontally and, once mounted, the beamline is fixed. Therefore the other 
sub-systems of LAUE apparatus must be aligned with respect to the beamline. The petal frame has been placed within the beamline 
cross section (in the Y-Z plane) and perpendicularly to its direction (X-axis). Also the source and the collimator slit have 
to move within the same cross section.  
Finally, the detectors were placed along the X-axis at 20 m from the petal frame. The structure 
holding the detectors was designed in such a way that both direct and  diffracted beam can be observable.

\begin{figure}[!h]
\begin{center}
\includegraphics[scale=0.26]{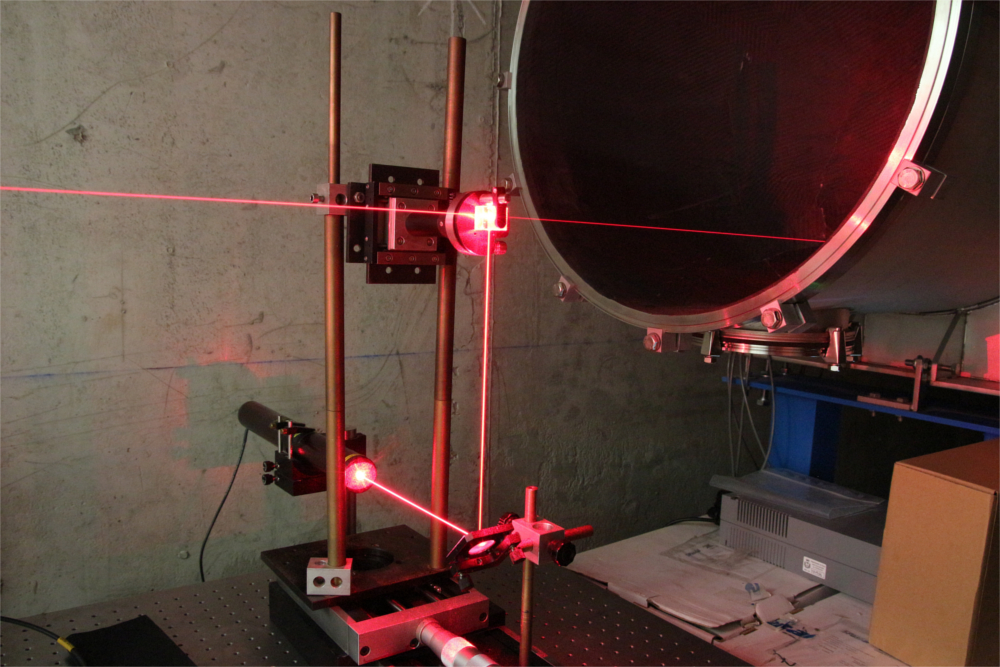}
\caption{\footnotesize The laser adopted to ensure the optical alignment.}
\label{fig:laser} 
\end{center}
\end{figure}

To ensure that all the translations of the source and  collimator slit (in the Y-Z plane) are kept parallel, a further optical phase of 
alignment is adopted. The method is based on a laser, mounted on a optical bench set between the beam-line 
and the motorized collimator slit, that emits an horizontal light beam. The  beam is reflected by a mirror 
placed at 45$^o$, then is sent to a beam-splitter which splits the beam along opposite directions, one towards the hexapod crystal holder, positioned in a reference lens frame position, through the collimator slit and the other toward the gamma-ray source through the beamline (see Fig.~\ref{fig:laser}). Mirrors placed over the gamma-ray source and over the crystal holder 
reflect back the two beams. Their superposition in the laser output  
ensures the correct zero-position for source,  collimator, hexapod holding aperture, and the petal reference position.

The optical alignment ensures that the ideal line connecting the source and the collimator slit is
parallel to the lens axis. It also guarantees an excellent precision in positioning the center of the 
gamma--ray imager in correspondence of the lens focus.

\begin{figure}[!h]
\begin{center}
\includegraphics[scale=0.21]{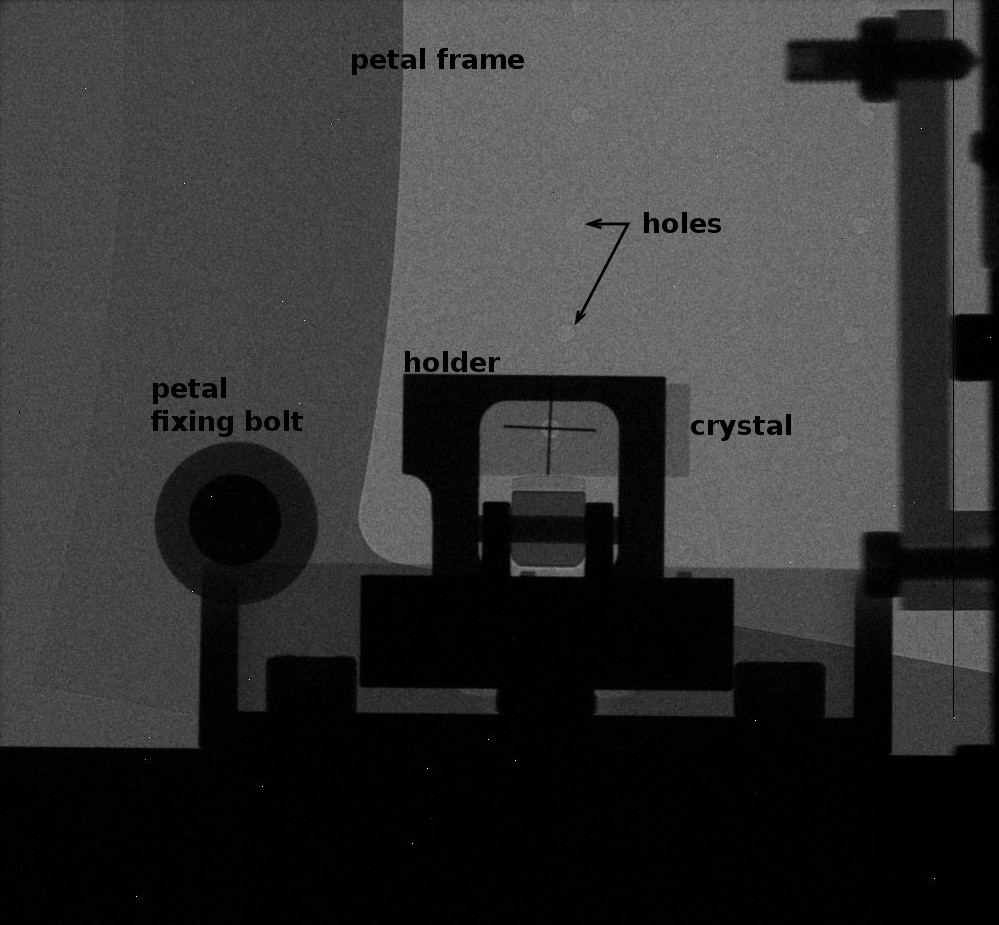}
\caption{\footnotesize The radiography of the entire apparatus once the collimator has been shifted out 
of the X-ray beam. Part of the petal, the holder and a dummy crystal with the tungsten cross are visible.}
\label{fig:radiography} 
\end{center}
\end{figure} 

\begin{figure}[!h]
\begin{center}
\includegraphics[scale=1.1]{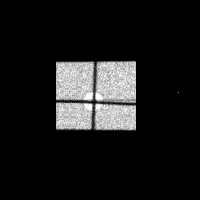}
\includegraphics[scale=1.1]{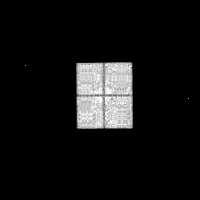}
\caption{\footnotesize Images obtained before and after the fine tuning of the alignment process. Initially the two 
Tungsten crosses placed on the collimator slit and on the crystal holder were slightly misaligned ({\it left})
while after a shift and a tilt of the X-ray source carriage, the crosses result superposed ({\it right}).}
\label{fig:crosses} 
\end{center}
\end{figure}

The gamma-ray alignment performs a fine tuning of the previous alignment process. It has been
performed using two Tungsten crosses (wire diameter of 200 $\mu$m) that are placed respectively on the center of a dummy crystal positioned  
in the crystal holder, and on the aperture center of the slit.
The crystal holder, by means of the hexapod, was ina position such that the tungsten cross center was placed in correspondence of the frame hole from which the resin is injected to fix the crystal to the frame. 
In Fig.~\ref{fig:crosses} we show the crosses before and after the fine tuning.

\section{First results}

\begin{figure}[!h]
\begin{center}
\includegraphics[scale=1.35]{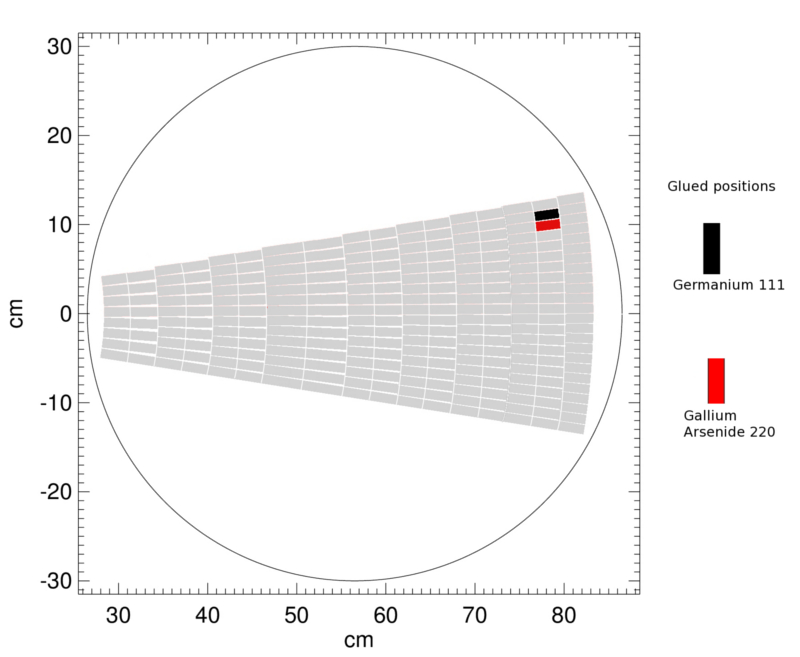}
\caption{\footnotesize Demonstrative model with indicated where the two crystals of Ge (111) and GaAs (220) were placed.}
\label{fig:petalodimostrativo} 
\end{center}
\end{figure}

To test the apparatus correct alignment, we performed the crystal orientation of two crystal tiles (one of Ge (111) 
and the other of GaAs (220)) and then, using the adopted resin, we fixed them to the petal frame. For this test, the 
two crystals were placed one next the other (cells \#24 and \#25). In this way we could also check the capability of 
the hexapod to position them with the required inter-space of 0.2 mm. A sketch of the whole set of crystal cells and those 
two selected  is shown in Fig.~\ref{fig:petalodimostrativo}.

The first results were also exploted to compare expected and measured energy of the diffracted beam in correspondence of  
chosen crystal cell. We also tested orientation stability of the tiles  once the resin had polimerized.

\subsection{Spectral results}

The right orientation of each crystal on the frame is achieved when the beam diffracted by the crystal has a peak energy 
corresponding to the energy extected for the selected cell. A Ge (111) tile was first tested. Its orientation was that 
expected for the cell \#24 with nominal centroid energy of 96.14 keV. 

\begin{figure}[!h]
\begin{center}
\includegraphics[scale=0.72]{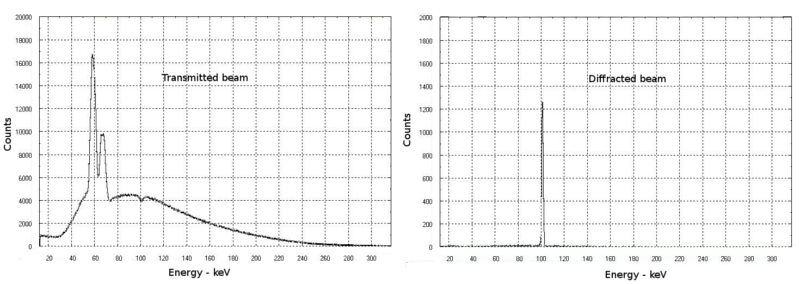}
\caption{\footnotesize Spectrum of the beam through a Ge (111) tile, when it is oriented for diffraction. {\it Left}: transmitted spectrum. {\em Right}: diffracted spectrum. Exposure time = 300 sec. 
The beam pass through a collimator slit whose size is 0.5 $\times$ 6 mm$^2$.}
\label{fig:ge_td} 
\end{center}
\end{figure}

\begin{figure}[!h]
\begin{center}
\includegraphics[scale=0.72]{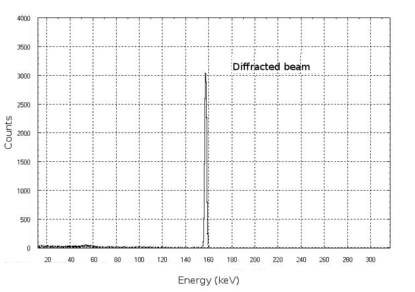}
\includegraphics[scale=0.72]{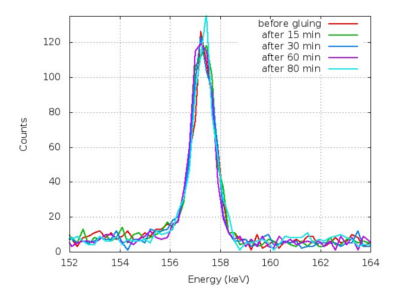}
\caption{\footnotesize {\em Left}: Spectrum of the diffracted photons from the GaAs (220) tile placed in the cell \#25. 
The nominal energy for its position is 157.10 keV. Exposure time = 300 sec. 
The beam pass through the collimator slit whose size is 0.5 $\times$ 6 mm$^2$. 
{\em Right}: After the injection of the adhesive, the GaAs (220) has been systematically monitored 
during the 80 minutes of the polimerization phase. No significant deviation with respect to the nominal energy
was observed.}
\label{fig:gaas_d} 
\end{center}
\end{figure}

In Fig.~\ref{fig:ge_td}~{\it left} it is shown the transmitted 
beam through the crystal when the proper diffraction angle was set. A dip at about 100 keV is clearly visible. When the 
spectrometer is translated to the lens focus,  the diffracted spectrum is shown in Fig.~\ref{fig:ge_td}~{\it right}.    
Similarly, for the GaAs (220) tile placed  in the contiguous cell,  the spectrum of the diffracted beam is shown
in Fig.~\ref{fig:gaas_d}~{\it left}.

A systematic monitoring of the diffracted beam was performed during the gluing procedure. The spectrum was 
acquired every 15 minutes after the adhesive injection. Figure~\ref{fig:gaas_d}~{\it right} shows that the effect of the 
glue polimerization does not produce any significant deviation of the centroid energy with respect to the nominal 
energy expected for the  GaAs (220) placed in the cell \#25.

\begin{figure}[!h]
\begin{center}
\includegraphics[scale=0.32]{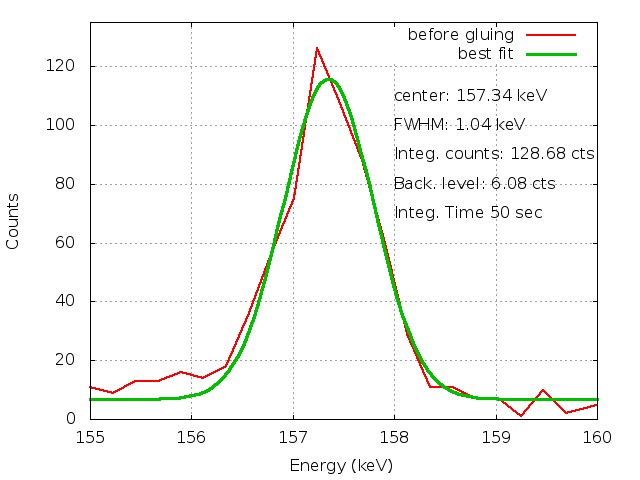}
\includegraphics[scale=0.32]{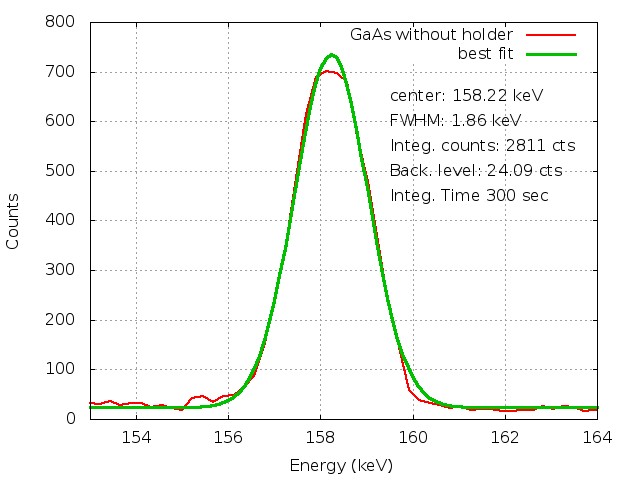}
\caption{\footnotesize Diffracted spectra of the GaAs (220)before and after the mechanical 
release of the holder. The difference is visible in terms of energy deviation from the nominal position. After the holding removal 
the energy deviation was within 1 keV, corresponding to $\sim$ 20 arcseconds. The effect is known and mainly due to a not uniform 
glue distribution between the crystal and the carbon fiber support, given the angular tilt applied to the crystal. This effect 
can be taken into account during the assembling phase.}
\label{fig:gaasprimadopo} 
\end{center}
\end{figure}

After the glue polimerization, the holder removal is the most critical phase. 
With the GaAs (220) the centroid energy changed by less than 1 keV, corresponding to less that $\sim$ 20 arcsec. 
Figure~\ref{fig:gaasprimadopo} shows
the measured spectra before and after the holding removal. The effect is mainly related to the angular tilt between
the petal frame and the glued crystal plane. The tilt produces an asymmetric glue distribution 
over the crystal, giving origin to the observed energy change. The same effect was also observed during the preliminary  resin
tests in which the orientation of 20 muck-up tiles of 30 $\times$ 10 mm$^2$ cross section was monitored during 
their polimerization phase. A systematic deviation of 20-30 arcseconds was found.
This systematic error is being taken into account during the assembling phase.      

\newpage
\subsection{Images of the diffracted beam}

The diffracted beam was also imaged by setting the center of the imager (pixels $(x, y)= (512, 512)$) in the nominal lens focus. 
Due to the beam divergence, it is not possible to hit the entire crystal and observe the complete focusing effect from it. 
Instead, it is possible to observe the focusing capability when the beam size radiates only the central portion of the tile, making the beam divergence negligible (few arcesconds). Therefore the slit aperture was set to  4 $\times$ 6 mm$^2$ with the smaller size radially.

\begin{figure}[!h]
\begin{center}
\includegraphics[scale=0.35]{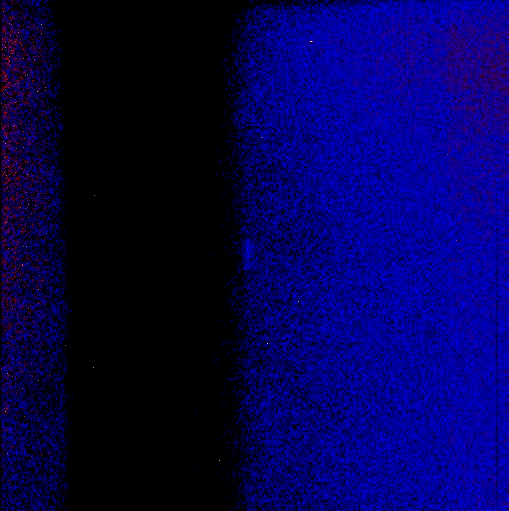}
\caption{\footnotesize Image of the diffracted beam from the GaAs (220) bent crystal. The beam cross section 
is 4 $\times$ 6 mm$^2$. Being the image size 2 $\times$ 12 mm$^2$, the focusing effect along the horizontal 
dimension is visible. On the other hand, the diffracted vertical dimension is bigger than the slit size for 
divergence effects.}
\label{fig:diffgaas} 
\end{center}
\end{figure}

The focusing effect can be observed in Fig.~\ref{fig:diffgaas} in which the radial dimension has decreased from 
4 mm in correspondence of the crystal to 2 mm at the detector level due to the focusing capability of the crystal. 
In the orthogonal direction, non focusing effect is expected and, indeed, in this case the beam 
increases from 6 mm to 12 mm as expected due to the beam divergence. 

\section{Conclusions}

We have described the LAUE project supported by the Italian Space Agency (ASI), in particular the facility developed, in the LARIX laboratory of the University of Ferrara, for assembling and testing Laue lens petals for space applications.
The expected accuracy in the lens assembling would allow to build
lenses even with very long focal lengths (up to 100 m), a goal never achieved so far. Removing possible 
systematic effects from the glue polimerization, the goal of 10 arcseconds accuracy appears to be satisfied.

In addition to lens assembling technology the LAUE project has faced the crystal production of proper crystals for 
Laue lenses. For the first time we have developed and tested bent crystals. They appear the most suitable for their high 
diffraction efficiency and focusing capability.

As a demonstration of the validity of the adopted technology, a lens petal of 20 m focal length, made of $\sim$ 300 bent 
crystal tiles of Ge(111)  and GaAs(220) is being assembled. We have reported preliminary results on the firsts glued crystals.

\bibliography{bibliography}
\bibliographystyle{spiebib}

\end{document}